\documentclass[conference]{IEEEtran}
\IEEEoverridecommandlockouts
\usepackage{cite}
\usepackage{amsmath,amssymb,amsfonts}
\usepackage{algorithmic}
\usepackage{siunitx}
\usepackage{graphicx}
\usepackage{textcomp}
\usepackage{xcolor}
\usepackage[a4paper, total={184mm,239mm}]{geometry}
\def\BibTeX{{\rm B\kern-.05em{\sc i\kern-.025em b}\kern-.08em
        T\kern-.1667em\lower.7ex\hbox{E}\kern-.125emX}}

\usepackage{fancyhdr}
\usepackage[all]{background}
\usepackage{stackengine}
\setstackEOL{\\}
\setstackgap{L}{\normalbaselineskip}
\SetBgContents{\color{gray}{\tiny \Longstack{PREPRINT - accepted In Proceedings of the 25th Conference on Design, Automation and Test in Europe (DATE '22)}}}
\SetBgPosition{4.2cm,1cm}
\SetBgOpacity{1.0}
\SetBgAngle{0}
\SetBgScale{1.8}

\usepackage{pgfplots}
\usepackage{pgfplotstable}
\usetikzlibrary{patterns, arrows, pgfplots.groupplots,hobby}
\usetikzlibrary{patterns}
\usetikzlibrary{positioning}
\usetikzlibrary{pgfplots.groupplots}
\usetikzlibrary{plotmarks}
\usetikzlibrary{calc}
\usepgfplotslibrary{external}
\usepackage{balance}
\usepackage{caption}
\usepackage{bm}
\usepackage{subfig}
\usepackage{enumitem}
\usepackage{soul}
\definecolor{cg1}{HTML}{E8F1FA}
\definecolor{cg2}{HTML}{C7DDF2}
\definecolor{cg3}{HTML}{8EBAE5}
\definecolor{cg4}{HTML}{407FB7}
\definecolor{cg5}{HTML}{00549F}
\definecolor{red}{HTML}{CC071E}

\usepackage{keyval}

\makeatletter
\newcommand\Row[1]{%
    \par\nobreak\nointerlineskip\vskip-\fboxrule%
    \@tfor\@tempa:=#1 \do {\csname ChessBox\@tempa\endcsname\kern-\fboxrule}}
\define@key{chessB}{side}{\def\ChessSide{#1}}
\define@key{chessB}{colori}{\def\ChessColori{#1}}
\define@key{chessB}{colorii}{\def\ChessColorii{#1}}
\define@key{chessB}{coloriii}{\def\ChessColoriii{#1}}
\setkeys{chessB}{
    side=1.5em,
    colori=black!70,
    colorii=white,
    coloriii=black}
\makeatother

\renewcommand*\descriptionlabel[1]{\hspace\leftmargin$#1$}

\begin{document}
\bstctlcite{IEEEexample:BSTcontrol}

\title{NeuroHammer: Inducing Bit-Flips in Memristive Crossbar Memories
\thanks{This work was funded by the Federal Ministry of Education and Research (BMBF, Germany) in the project NEUROTEC II (Project Nos. 16ME0398K and 16ME0399).}
\vspace{-4mm}
}

\author{
    \IEEEauthorblockN{Felix Staudigl\IEEEauthorrefmark{1},
                                     Hazem Al Indari\IEEEauthorrefmark{1},
                                     Daniel Sch\"on\IEEEauthorrefmark{3},
                                     Dominik Sisejkovic\IEEEauthorrefmark{1},\\
                                     Farhad Merchant\IEEEauthorrefmark{1},
                                     Jan Moritz Joseph\IEEEauthorrefmark{1}
                                     Vikas Rana\IEEEauthorrefmark{2}
                                     Stephan Menzel\IEEEauthorrefmark{3},
                                     Rainer Leupers\IEEEauthorrefmark{1}
                             }
    \IEEEauthorblockA{\IEEEauthorrefmark{1}
        \textit{Institute for Communication Technologies and Embedded Systems, RWTH Aachen University, Germany}
    }
    \IEEEauthorblockA{\IEEEauthorrefmark{2}
        \textit{Peter Gr\"unberg Institut (PGI-10) Forschungszentrum Juelich GmbH, Juelich, Germany}
    }
    \IEEEauthorblockA{\IEEEauthorrefmark{3}
    \textit{Peter Gr\"unberg Institut (PGI-7) Forschungszentrum Juelich GmbH, Juelich, Germany}\\
    \{staudigl, alindari, sisejkovic, merchantf, joseph, leupers\}@ice.rwth-aachen.de\\
    \{st.menzel, schoen, v.rana\}@fz-juelich.de\\
}
\vspace{-8mm}
}

\maketitle

\begin{abstract}
Emerging non-volatile memory (NVM) technologies offer unique advantages in energy efficiency, latency, and features such as computing-in-memory. Consequently, emerging NVM technologies are considered an ideal substrate for computation and storage in future-generation neuromorphic platforms. These technologies need to be evaluated for fundamental reliability and security issues. In this paper, we present \emph{NeuroHammer}, a security threat in ReRAM crossbars caused by thermal crosstalk between memory cells. We demonstrate that bit-flips can be deliberately induced in ReRAM devices in a crossbar by systematically writing adjacent memory cells. A simulation flow is developed to evaluate NeuroHammer and the impact of physical parameters on the effectiveness of the attack. Finally, we discuss the security implications in the context of possible attack scenarios.
\end{abstract}


\section{Introduction}
\begin{figure*}[ht]
    \centering
    \includegraphics[width=\textwidth]{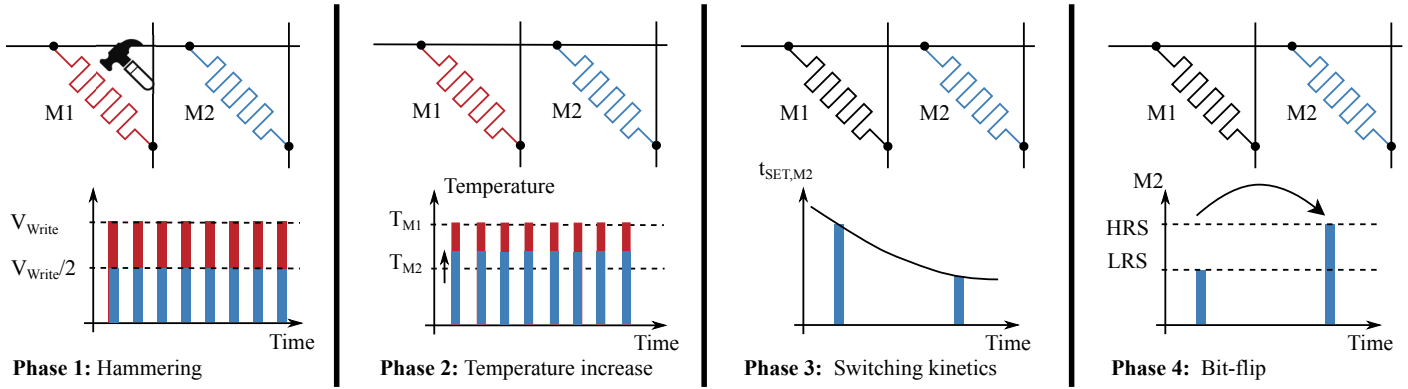}
    \caption{Working principle of NeuroHammer.}
    \label{fig:neuro_hammer_mechanis}
    \vspace{-5mm}
\end{figure*}
Emerging non-volatile memory (NVM) technologies offer promising features that give them an advantage over classical RAM technologies. These advantages are density, low leakage power, and computing-in-memory (CIM) capabilities~\cite{neurosurvey1}. Specifically,  CIM capabilities help alleviate the von Neumann bottleneck by significantly reducing data movements~\cite{Staudigl_2021}.

Reliability and disturbance errors in contemporary solid-state storage technologies, like dynamic random access memory (DRAM), have emerged due to a technology push resulting in high memory densities~\cite{rowhammer1,Kim_2014}. These disturbance errors can form security vulnerabilities like RowHammer, compromising user information through privilege escalation, as demonstrated in Google's Project Zero~\cite{Seaborn2015}, or denial of service in clouds~\cite{rowhammer4}.

Similar disturbance errors have been reported in phase-change memory, which suffers from thermal crosstalk in technology nodes below \SI{20}{\nm}~\cite{Jiang2014}. Cai et al.~\cite{Cai2020} discuss thermal crosstalk in ReRAM structures and their impact on the reliability of neuromorphic systems. These disturbance errors can cause an unintended malfunction in memory cells. The authors in~\cite{Witzleben2017} discuss the impact of the filament temperature on the switching kinetics of ReRAM cells. In this paper, we combine the reported thermal crosstalk in dense memristive crossbars with the temperature impact on the switching kinetics to define an attack that intentionally causes malfunctions.

\textbf{Contributions:}~We introduce \textit{NeuroHammer}, a security attack on emerging NVMs, which deliberately causes bit-flips. To the best of our knowledge, this paper provides the first investigation of undesired bit-flips in emerging NVMs in the context of a security attack. The major contributions are as follows. (1) A security threat based on disturbance errors in ReRAM structures through thermal crosstalk observed during the writing process of memory cells. (2) A simulation methodology for the characterization of NeuroHammer using state-of-the-art crossbar models. (3) A discussion on NeuroHammer-induced security threats based on the provided characterization.

\section{Background}\label{sec:back}

\textbf{RowHammer:}~In 2014, Kim et al.~\cite{Kim_2014} reported the impact of disturbance errors on DRAMs. The DRAM memory cell consists of a capacitor connected to a transistor. The capacitor's charge decreases with the scaling down of the DRAM process technology. Consequently, the memory cell is more susceptible to disturbance errors based on electromagnetic inference. The RowHammer attack uses this phenomenon to deliberately flip certain bits in DRAM memories by hammering/reading particular rows. The consciously triggered bit-flips violate a fundamental concept of secure and reliable computing systems: memory isolation, which ensures strict separation of application memory to mitigate malicious changes in its internal state.

\textbf{ReRAM}:~Redox-based resistive memories (ReRAMs) are an emerging class of non-volatile memories, which store binary information in terms of resistances, i.e., the \textit{low resistive state} (LRS) and the \textit{high resistive state} (HRS). ReRAMs typically consist of a simple metal/insulator/metal structure, enabling a high integration density.
Depending on the ionic defect type, one can distinguish between valence change memories (VCM) and electrochemical metallization cells (ECM) \cite{Waser2009}. VCM cells rely on the motion of oxygen defects in the insulating oxide material. ECM cells are based on the migration of (typically) Ag or Cu cationic defects, which were injected from the chemically active electrode consisting of Ag or Cu.
The most studied ReRAMs are filamentary VCM cells in which oxygen defects form very small conducting filamentary regions. This enables high scalability down to a few nm\cite{Pi2019} and fast switching speed down to 50 ps \cite{Witzleben2020} at moderate switching voltages. This fast switching speed is achieved by local Joule heating, which accelerates the ion migration exponentially \cite{Menzel2011}.

\section{Mechanics of NeuroHammer}
\label{sec:mechanics}
\label{sec:neurohammer}
\begin{figure*}[ht]
    \centering
    \subfloat[]{
        \includegraphics[width=0.23\textwidth]{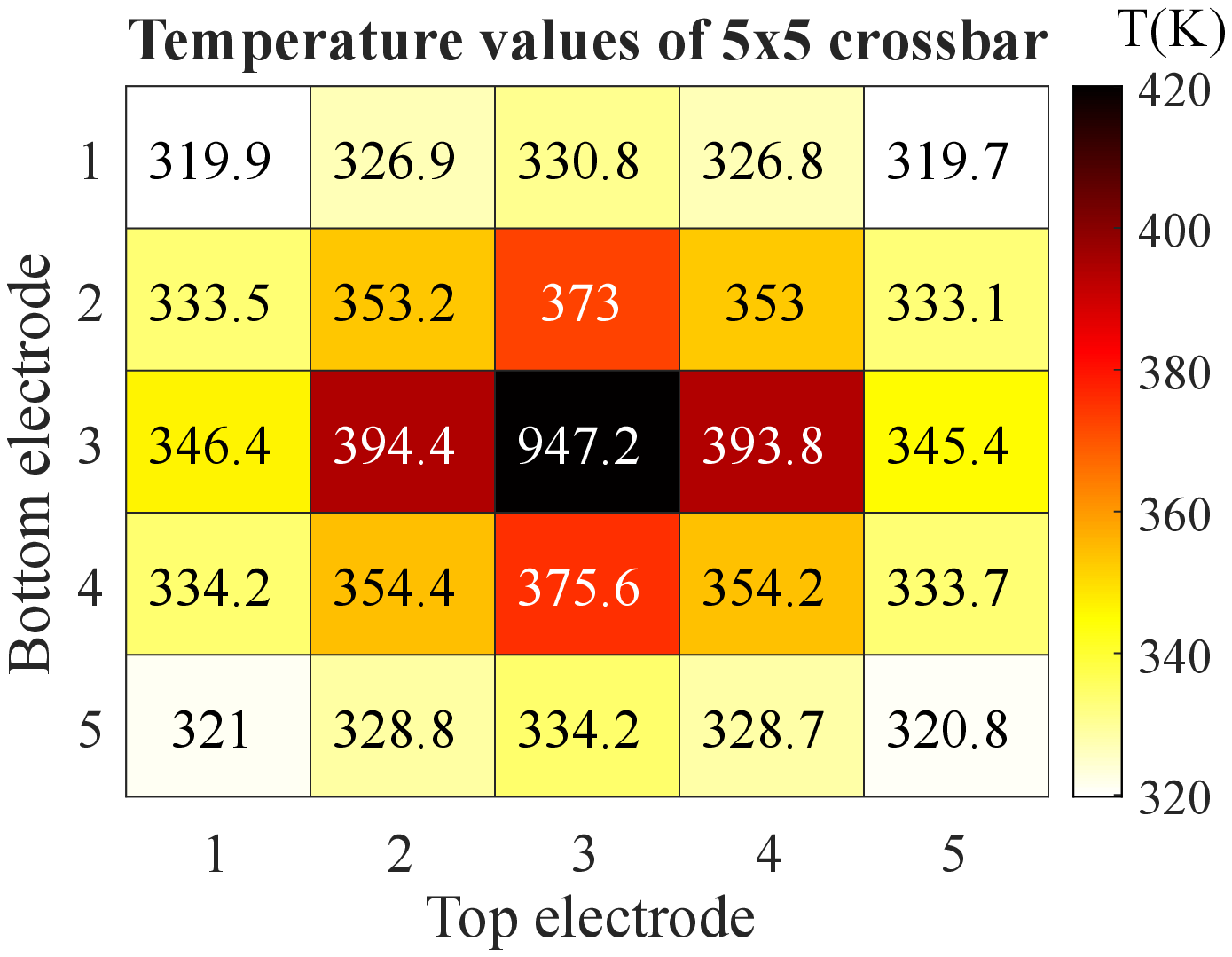}
        \label{fig:heatmap_alphas}
    }\subfloat[]{
        \includegraphics[width=0.32\textwidth]{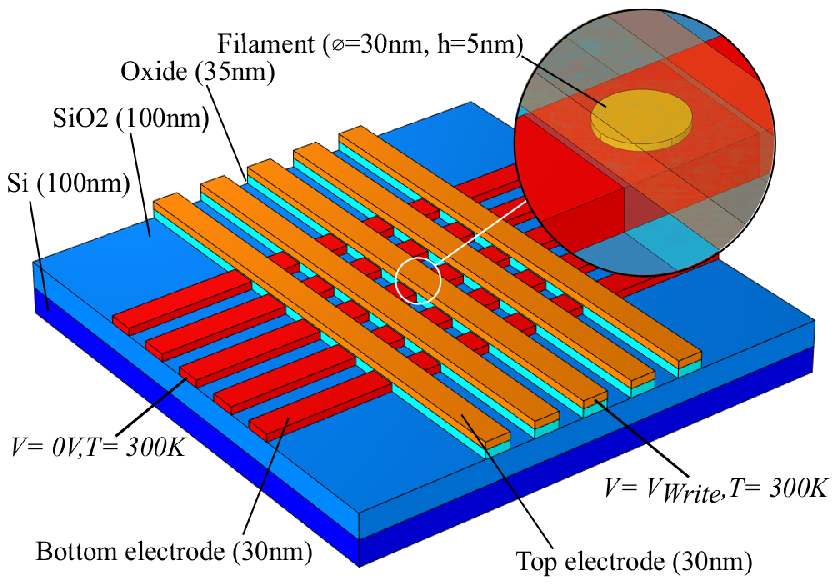}
        \label{fig:device_level_simulation_comsol}
    } \hfil \subfloat[]{
        \includegraphics[width=0.23\textwidth]{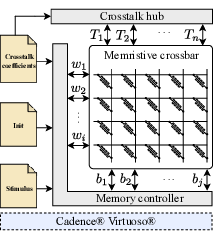}
        \label{fig:simu_framework_overview}
    }
    \vspace{-2mm}
    \caption{Simulation methodology: (a) the thermal coupling in a 5x5 memristive crossbar, (b) low-level simulation setup and its boundary conditions, and (c) circuit-level simulation setup and its internal components.}
    \vspace{-5mm}
\end{figure*}
Von Witzleben et al.~\cite{Witzleben2017} investigated the impact of the temperature on the switching kinetics of redox-based ReRAM cells, concluding that the switching time depends significantly on the temperature of the filament. The proposed attack uses this phenomenon to deliberately induce bit-flips in ReRAM crossbar memories. In the following, we define a \textit{pulse} as a rectangular electrical pulse with a fixed amplitude of $V_\text{SET} = 1.05\text{V}$ and a given pulse length. The pulse length is defined by the duration in which the signal is active. The attack is divided into four distinct phases, as illustrated in Fig.~\ref{fig:neuro_hammer_mechanis}:

\subsubsection{Hammering} The NeuroHammer attack hammers a single memory cell to trigger a bit-flip. In Fig.~\ref{fig:neuro_hammer_mechanis}, the red (blue) cell symbolizes the attacked (target) cell. The red cell should be initially switched to LRS to maximize the resulting current flowing through the cell. The $V/2$ scheme is used to apply constant stress to the blue cell, which equals $V_\text{SET}/2$ of the voltage pulse applied to the red cell.

\subsubsection{Temperature increase} The recurring voltage pulses applied on the red cell result in a current through the device, and hence, a temporary temperature increase of the filament. As a result, the temperature of the blue cell increases due to thermal coupling. In addition, the $V/2$ scheme applies a respective voltage pulse ($V_\text{SET}/2$) to the blue cell, which simultaneously leads to a temperature increase in the filament.

\subsubsection{Switching kinetics} The increased temperature of the blue cell changes the switching kinetics of the ReRAM device. Consequently, the ReRAM device is more susceptible to pulses, which have either a larger amplitude or a longer pulse length.

\subsubsection{Bit-flip} At this point, the blue cell gradually switches its internal state based on the changed switching kinetics and the constantly applied $V/2$ voltage pulses. These pulses would typically not be sufficient to switch the state of the blue cell. However, considering the thermal coupling effects of dense crossbar structures, the target cell is susceptible to pulses with even lower amplitudes. Eventually, the blue cell switches its internal state and a bit-flip occurs.

\section{Simulation Methodology}
\label{sec:methodology}
To verify the attack presented in Chapter~\ref{sec:mechanics}, a simulation framework was developed based on the simulator COMSOL Multiphysics~\cite{comsol} and the circuit simulator Cadence Virtuoso~\cite{virtuoso}. Initially, the crossbar simulation aims to simulate the thermal crosstalk, based on which we extract thermal crosstalk coefficients, which are called \textit{alpha values}. These values serve as input for the circuit simulation to enable the framework to simulate thermal coupling in dense ReRAM crossbar structure.

\subsection{Crossbar Simulation}
\label{subsec:device_level_simulation}
Fig.~\ref{fig:device_level_simulation_comsol} illustrates the basic structure of the simulation model, which consists of a memristive crossbar array of electrodes on a Si/SiO$_2$ substrate. To calculate the temperature of the adjacent cells which surround the target cell, the static heat transfer equation and the current continuity equation are solved for the temperature \(T\) and the potential \(\phi\), shown in Equation \ref{eq:static_heat} and \ref{eq:current}.
\begin{equation}\label{eq:static_heat}
    -\nabla\cdot(\kappa\nabla T) = \textbf{j}\cdot\textbf{E}
\end{equation}
\begin{equation}\label{eq:current}
    \nabla\cdot\textbf{j} = -\nabla\cdot(\sigma\nabla\phi) = 0
\end{equation}
Here, \(\kappa\) denotes the thermal conductivity, \(\sigma\) the electric conductivity, \(\textbf{j}\) the local current density, and \(\textbf{E}\) the electric field. The electric conductivity and thus also the thermal conductivity (see Wiedemann–Franz law) of the filament is adjusted so that a certain current flows through the device. All other surfaces are thermally and electrically insulated.

To determine the alpha values of a geometry with specific materials for a temperature prediction of the adjacent cells, a variation of the dissipated power in the selected cell is required. Through a voltage sweep of \(V_\text{SET}\), a temperature matrix which contains the temperature of each cell can be extracted from every simulation (see Fig. \ref{fig:heatmap_alphas}). The thermal resistance \(R_\text{th}\) of the selected cell can then be determined as a fit parameter of a linear regression of the respective temperature \(T(P_\text{LRS})\) and dissipated power \(P_\text{LRS}=V_\text{SET}\cdot I\) (Equation \ref{eq:linear_reg_rth}). With further linear regression for every neighboring cell, the alpha values can be determined (Equation \ref{eq:linear_reg_alpha}).
\begin{equation}\label{eq:linear_reg_rth}
    T(P_\text{LRS}) = T_0 + R_\text{th} \cdot P_\text{LRS}
\end{equation}
\begin{equation}\label{eq:linear_reg_alpha}
    T_{ij}(P_\text{LRS}) = T_0 + R_{th} \cdot P_\text{LRS} \cdot \alpha_{ij}
\end{equation}
Here, \(T_{ij}(P_\text{LRS})\) denotes the temperature of the cell \(i,j\) depending on the dissipated power \(P_\text{LRS}\) and \(\alpha_{ij}\), the alpha value of the cell \(i,j\).

\subsection{Circuit Simulation}
The circuit-level simulation and its internal modules are illustrated in Fig.~\ref{fig:simu_framework_overview}. The framework consists of three major parts: memristive crossbar, memory controller, and the crosstalk hub. The platform can be parameterized based on configuration files and the standard graphical user interface (GUI) of the Cadence Virtuoso tool. The remainder of this section describes the modules of the framework in detail.

\textbf{Memory controller:}~In general, crossbar structures are interfaced over their rows (word lines) and columns (bit lines). The memory controller is responsible for generating and driving the respective pulse for a certain input line of the crossbar. The stimuli file stores the explicit characteristics, i.e. pulse length, duty cycle, and amplitude of each input pulse, while the init file holds the initial state of every ReRAM cell.

\textbf{Crosstalk hub:}~The hub calculates the temperature increase of a ReRAM cell due to thermal coupling with all surrounding cells. The temperature is calculated based on the alpha values extracted from the crossbar simulation (Section~\ref{subsec:device_level_simulation}) and the filament temperatures of the adjacent cells:
\begin{equation}
    T_\text{in}(\bm{\alpha},\textbf{T}) = \sum_{\substack{
            0<i<m \\
            0<j<n
    }}
    \alpha_{ij} T_{ij,\text{out}},
\end{equation}
where~\(i,j \in \mathbb{N}\) indicate the row and column of the crossbar excluding the attacked cell itself,~\(m,n \in \mathbb{N}\) indicate the number of rows and columns, and~\(T_\text{in}\) the additional temperature based on the thermal crosstalk.

\textbf{Memristive crossbar:}
The passive 5x5 memristive crossbar array represents the central part of the simulation framework in which the instantiated memristive devices connect the bit lines to the respective word lines. The JART VCM v1b model employed in this study was developed for filamentary switching VCM cells and fitted to a nanocrossbar Pt/HfO$_2$/TiO$_\text{x}$/Ti device \cite{Cueppers2019,Bengel2020,JART2019}. The dissipated power $P_\text{d}$ in the cell increases the local temperature $T$ according to
\begin{equation}
    T = R_\text{th,eff} \cdot P_d + T_0,
\end{equation}
where $T_0$ is the ambient temperature and $R_\text{th,eff}$ is the effective thermal resistance (in K/W), describing the heat dissipation to the immediate cell surrounding and the thermal properties of the materials. The complete equation system and the used parameters can be found in~\cite{Bengel2020}. It should be noted that the "deterministic" model version is used here.

The original model was adjusted to enable the memristive model to exchange parameters with the simulation framework. Thus, we introduced two interface variables to communicate the temperature values to the crosstalk hub, and receive the additional temperature generated from the adjacent cells.

\section{Results}
\label{sec:res}
\begin{figure*}[ht]
    \centering
    \subfloat[]{
        \begin{tikzpicture}[scale=0.5]
            \pgfplotstableread[row sep=\\,col sep=&]{
                time &  v1 \\
                10   & 5530 \\
                15   & 3281 \\
                20   & 2466 \\
                30   & 1564 \\
                45   &1064 \\
                50   & 943 \\
                60   & 797 \\
                75   & 629 \\
                100 & 472 \\
            }\dataset
            \begin{semilogyaxis}[
                ylabel={\# pulses to trigger a bit-flip},
                xlabel={Pulse length in~\SI{}{\ns}},
                ylabel style={at={(-0.004,0.5)},anchor=north},
                width=9cm,
                height=6cm,
                ymin=0,
                ymax=10000,
                ytick={0, 1, 10, 100, 1000, 10000},
                xtick={10, 20, 30, 40, 50, 60, 70, 80, 90, 100}, 
                ylabel style ={font=\Large},
                xlabel style ={font=\Large},
                tick label style={font=\large},
                ]
                \addplot[draw=cg5,mark=*] table[x=time,y=v1] \dataset;
            \end{semilogyaxis}
        \end{tikzpicture}
        \label{fig:sim_pulse_length}
    } \hfil \subfloat[]{
        \begin{tikzpicture}[scale=0.5]
            \pgfplotstableread[row sep=\\,col sep=&]{
                time & v1 & v2 & v3\\
                10 & 1594 &	1062 &	797 \\
                50 & 6946 &	4629 &	3474 \\
                90 & 31294 &	20849 &	15648\\
            }\dataset
            \begin{axis}[ybar,
                ylabel={\# pulses to trigger a bit-flip},
                xlabel={Electrode spacing},
                ylabel style={at={(-0.01,0.5)},anchor=north},
                legend style={at={(0.5,0.96), font=\Large},
                    anchor=north,legend cell align={left}},
                legend image post style={scale=1.4},
                enlarge x limits={abs=1.1cm},
                ymajorgrids = true,
                bar width=.5cm,
                ymode = log,
                width=9.5cm, height=6cm,
                ymin=0,ymax=100000,
                ytick={0, 1, 10, 100, 1000, 10000, 100000},
                xtick align=inside,
                xtick=data,
                xticklabel style={align=center},
                xticklabels = {
                    \SI{10}{\nm},
                    \SI{50}{\nm},
                    \SI{90}{\nm}
                },
                ylabel style ={font=\Large},
                xlabel style ={font=\Large},
                tick label style={font=\large},
                legend entries={\SI{50}{\ns}, \SI{75}{\ns} , \SI{100}{\ns}},
                legend columns=6,
                ]
                \addplot[draw=black,fill=cg2] table[x=time,y=v1] \dataset;
                \addplot[draw=black,fill=cg3] table[x=time,y=v2] \dataset;
                \addplot[draw=black,fill=cg4] table[x=time,y=v3] \dataset;
            \end{axis}
        \end{tikzpicture}
        \label{fig:sim_spacing}
    }\hfil  \subfloat[]{
        \begin{tikzpicture}[scale=0.49]
            \pgfplotstableread[row sep=\\,col sep=&]{
                temp & v1 & v2 & v3 \\
                0 &    33030&	10735 &	6352 \\
                25 & 7139 &	2311 &	1367\\
                50 & 1818 &	589	& 349	\\
                75 & 539 & 175 & 104 \\
                100 & 184 & 60 & 35 \\
            }\dataset

            \begin{axis}[ybar,
                ylabel={\# pulses to trigger a bit-flip},
                xlabel={Ambient temperature},
                ylabel style={at={(-0.004,0.5)},anchor=north},
                legend style={at={(0.5,0.96), font=\Large},
                    anchor=north,legend cell align={left}},
                legend image post style={scale=1.4},
                enlarge x limits={abs=1.2cm},
                ymajorgrids = true,
                bar width=.5cm,
                ymode = log,
                width=15cm, height=6cm,
                ymin=0,ymax=100000,
                ytick={0, 1, 10, 100, 1000, 10000, 100000},
                xtick align=inside,
                xtick=data,
                xticklabel style={align=center},
                xticklabels = {
                    \SI{273}{\kelvin},
                    \SI{298}{\kelvin},
                    \SI{323}{\kelvin},
                    \SI{348}{\kelvin},
                    \SI{373}{\kelvin}
                },
                ylabel style ={font=\Large},
                xlabel style ={font=\Large},
                tick label style={font=\large},
                legend entries={\SI{10}{\ns}, \SI{30}{\ns} , \SI{50}{\ns}},
                legend columns=6,
                ]
                \addplot[draw=black,fill=cg2] table[x=temp,y=v1] \dataset;
                \addplot[draw=black,fill=cg3] table[x=temp,y=v2] \dataset;
                \addplot[draw=black,fill=cg4] table[x=temp,y=v3] \dataset;
            \end{axis}
            \label{fig:sim_ambient_temp}
        \end{tikzpicture}
    }

    \caption{Simulation results of the circuit-level simulation framework: (a) impact of the pulse length, (b) impact of the electrode spacing, (c) impact of the ambient temperature, (d) impact of different attack patterns, and (e-h) overview of attack patterns.}
\vspace{-5mm}
\end{figure*}
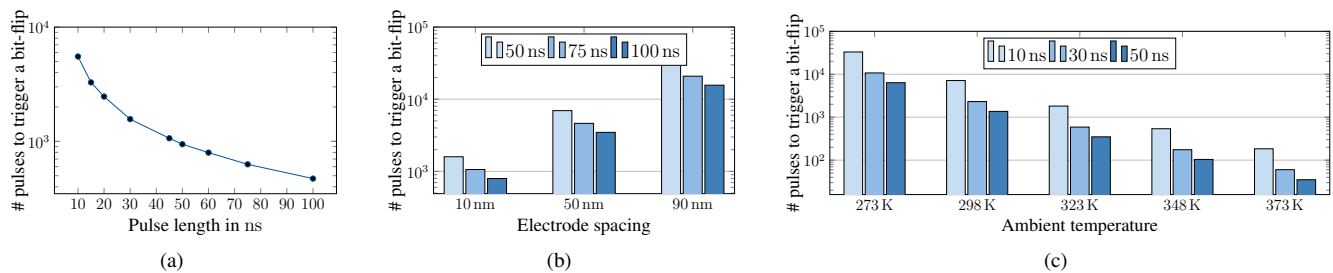

In this section, we verify the proposed NeuroHammer attack and demonstrate the flexibility of our simulation framework. Each experiment uses a simple attack pattern in which we attack the cell in the middle of the crossbar.

\textbf{Pulse length:}~The simulation uses a crossbar with an electrode spacing of \SI{50}{\nm}, assuming an ambient temperature of \SI{300}{\kelvin}. The controllers select the attacked cell by applying $V$ and $GND$. All remaining inputs are supplied with $V/2$ to minimize the sneak-path currents. Hence, the bit-flip can only occur in the blue cells based on the voltage drop of $V/2$ compared to the remaining cells, which experience no voltage drop. The results show that the number of required pulses to trigger the error decreases with the pulse length (Fig.~\ref{fig:sim_pulse_length}).

\textbf{Electrode spacing:}~The second experiment concerns the electrode spacing of the crossbar memory structure. The analysis uses an ambient temperature of \SI{300}{\kelvin}. The spacing of the electrode is defined by the distance between the electrodes of two adjacent cells. Fig.~\ref{fig:sim_spacing} illustrates the simulation results of the electrode spacing ranging from \SI{10}{\nm} to \SI{90}{\nm}, which indicate that the closer the cells are placed, the more vulnerable they are in terms of disturbance errors. Accordingly, we assume that disturbance errors become a serious problem for dense crossbar memory structures, the more the technology node advances.

\textbf{Ambient temperature:}~Next, we investigated the effects of the ambient temperature on the occurrence of disturbance errors. Fig.~\ref{fig:sim_ambient_temp} shows the results that use an electrode spacing of \SI{50}{\nm}. The results indicate a strong impact of the ambient temperature on the number of pulses required to trigger a bit-flip.


\section{Security Implications}
\label{sec:implications}
Several RowHammer-inspired attack scenarios have been presented to threaten both embedded systems and desktop computers. This section illustrates possible attack scenarios in which NeuroHammer can be exploited. Both attacks utilize disturbance errors in dense memory structures, which are deliberately caused by repeatedly hammering certain memory cells. Based on the similarities between both attacks, we briefly describe a common attack scenario using RowHammer to discuss, ultimately, the reuse of attack scenarios for NeuroHammer.

Seaborn et al.~\cite{Seaborn2015} demonstrated gaining kernel privileges through the DRAM RowHammer bug on a typical laptop. Two requirements enable the attack. First, the adversary has to find the correct address mapping between the physical and virtual memory space to hammer the correct cells. Second, the cache of the system would typically mitigate frequent activation of the same memory location. The authors used the flush instruction to constantly flush the cache, enabling the adversary to frequently activate the same memory location. Ultimately, the attacker gains kernel access to the whole physical memory through a normal process execution on an x86-64 machine by hammering a page table entry to point to an attacker-owned page table.

Considering its advantages, ReRAM has the potential to replace DRAM in modern computing systems. Thus, the mentioned attack scenario is transferable to NeuroHammer. Hereby, the access patterns must be adjusted to match the physical characteristics of the thermal crosstalk in memristive crossbar memories. In general, any attack proven to work with RowHammer could additionally work with NeuroHammer. However, the application domain of DRAM and ReRAM does not completely overlap. Hence, the proposed attack poses a supplementary threat to emerging neuromorphic-based systems, such as neuromorphic machine-learning accelerators.

\section{Conclusion}
\label{sec:con}
This work introduced NeuroHammer, a security threat to emerging non-volatile memories based on thermal crosstalk in dense crossbar structures. We developed a simulation framework to investigate the impact of physical parameters on the effectiveness of the attack. The results show that the NeuroHammer attack deliberately induces bit-flips in memristive crossbar structures. Finally, we discussed the security implications of NeuroHammer by analyzing attack scenarios of the related RowHammer attack. In future work, we plan to verify the existence of the NeuroHammer attack on physical crossbars and explore countermeasures to mitigate the security threat.

With this work, we demonstrated that the fundamental security implication of disturbance errors still persists even in emerging neuromorphic technologies. As neuromorphic hardware has the potential to become a key component of modern computing systems, evaluating its basic security aspects is essential to providing a stepping stone to building secure next-generation devices.

\vspace{-4mm}
\bibliographystyle{IEEEtran}
\bibliography{bibliography.bib}

\end{document}